\documentclass[twocolumn,showpacs,preprintnumbers,amsmath,amssymb,aps]{revtex4}

\usepackage{graphicx}% Include figure files
\usepackage{dcolumn}% Align table columns on decimal point
\usepackage{bm}% bold math

\begin{document}

\newcommand{\be}{\begin{eqnarray}}
\newcommand{\ee}{\end{eqnarray}}

\title{Anomalous Roughening in Experiments of Interfaces in Hele-Shaw Flows\\with Strong Quenched Disorder}

\author{Jordi Soriano}
\email{soriano@ecm.ub.es}

\author{Jordi Ort\'{\i}n}

\author{A. Hern\'andez-Machado}

\affiliation{Departament d'Estructura i Constituents de la Mat\`eria\\ Universitat de
Barcelona, Av. Diagonal, 647, E-08028 Barcelona, Spain}

\date{\today}

\begin{abstract}

We report experimental evidences of anomalous kinetic roughening in the stable displacement of
an oil--air interface in a Hele--Shaw cell with strong quenched disorder. The disorder
consists on a random modulation of the gap spacing transverse to the growth direction
(tracks). We have performed experiments varying average interface velocity and gap spacing,
and measured the scaling exponents. We have obtained $\beta \simeq 0.50$, $\beta^{\ast} \simeq
0.25$, $\alpha \simeq 1.0$, $\alpha_{loc} \simeq 0.5$, and $z \simeq 2$. When there is no
fluid injection, the interface is driven solely by capillary forces, and a higher value of
$\beta$ around $\beta=0.65$ is measured. The presence of multiscaling and the particular
morphology of the interfaces, characterized by high slopes that follow a L\'evy distribution,
confirms the existence of anomalous scaling. From a detailed study of the motion of the
oil--air interface we show that the anomaly is a consequence of different local velocities
over tracks plus the coupling in the motion between neighboring tracks. The anomaly disappears
at high interface velocities, weak capillary forces, or when the disorder is not sufficiently
persistent in the growth direction. We have also observed the absence of scaling when the
disorder is very strong or when a regular modulation of the gap spacing is introduced.
\end{abstract}

\pacs{47.55.Mh, 68.35.Ct, 05.40.-a}

\maketitle

\section{Introduction}\label{Sec:Introduction}

The study of the morphology and dynamics of rough surfaces and interfaces in disordered media
has been a subject of much interest in last years \cite{Barabasi-Stanley}, and is an active
field of research with relevance in technological applications and material characterization.
These surfaces or interfaces are usually self--affine, i.e. statistically invariant under an
anisotropic scale transformation. In many cases, the interfacial fluctuations follow the
dynamic scaling of Family--Vicsek (FV) \cite{Family-Vicsek-1985}, which allows a description
of the scaling properties of the interfacial fluctuations in terms of a growth exponent
$\beta$, a roughness exponent $\alpha$, and a dynamic exponent $z$. However, many growth
models \cite{Schroeder,krug,Das-Sarma,Dasgupta,Lopez-96,Castro}, experiments of fracture on
brittle materials, such as rock \cite{fracture-Granite-JuanMa} or wood
\cite{fracture-Wood-Engoy,fracture-Wood-Morel-1,fracture-Wood-Morel-2}, as well as experiments
on sputtering \cite{Sputtering-Jeffries}, molecular beam epitaxy \cite{MBE-Yang}, and
electrodeposition \cite{ED-Huo} have shown that the FV scaling is limited, and a new ansatz,
known as {\it anomalous scaling} \cite{Lopez-96,Lopez-97-I,Lopez-97-II}, must be introduced.
The particularity of the anomalous scaling is that local and global interfacial fluctuations
scale in a different way, so in general the global growth and roughness exponents, $\beta$ and
$\alpha$ respectively, differ from the local ones $\beta^{\ast}$ and $\alpha_{loc}$.

In experiments of fracture on brittle materials, while $\alpha$ is clearly dependent on the
material and the fracture mode, the measured exponents $\alpha_{loc}$ seem to have a universal
value, in the range $\alpha_{loc}^{3D}=0.8-0.9$ for three--dimensional fractures, and
$\alpha_{loc}^{2D}=0.6-0.7$ for two--dimensional fractures. L\'opez {\it et al.}
\cite{fracture-Granite-JuanMa} reported $\alpha = 1.2$ in an experiment of fracture in a
granite block, while in fracture of wood Morel {\it et al.} \cite{fracture-Wood-Morel-1}
obtained $\alpha=1.60$ and $\alpha=1.35$ for two different wood specimens. In these
experiments they also obtained $\beta=0.26$ and $\beta^{\ast}=0.14$. An interesting
observation from the experiments of fracture in wood, where tangential and radial fractures
were studied \cite{fracture-Wood-Engoy}, is that the morphology of the interfaces for the
former were characterized by large slopes, in contrast with the smoother appearance of the
latter. As pointed out by L\'opez {\it et al.} \cite{fracture-Granite-JuanMa}, the origin of
anomalous scaling in these experiments seems to be related with these large slopes in the
morphology of the interfaces.

In a previous work \cite{Soriano-2002-I} we presented an experimental study of forced
fluid invasion in a Hele-Shaw cell with quenched disorder. We focused on the limit of
weak capillary forces, and studied different types of disorder configurations at
different interface velocities and gap spacings. The results obtained showed that the
interfacial fluctuations scaled with time through a growth exponent $\beta \simeq 0.50$,
which was almost independent of the experimental parameters. The values of the roughness
exponents were characterized by two regimes, $\alpha_1$ at short length scales and
$\alpha_2$ at long length scales. For high interface velocities, we obtained $\alpha_1
\simeq 1.3$ and $\alpha_2 \simeq 0$, independently of the disorder configuration. For
moderate and low velocities, however, different values were obtained depending on the
disorder configuration, and different scaling regimes were reported. In addition, when
the disorder was totally persistent in the direction of growth, and for sufficiently
strong capillary forces, a new regime emerged. This new regime must be described in the
framework of anomalous scaling and was discussed in a recent Letter
\cite{Soriano-2002-PRL}. We obtained the scaling exponents for a specific set of
experimental parameters, and explored the range of validity of the anomaly. We also
introduced a phenomenological model that reproduces the interface dynamics observed
experimentally and the scaling exponents.

The objective of the present paper is to analyze the experimental evidences of anomalous
scaling in a systematic way, introducing additional techniques for the detection of anomalous
scaling, such as multiscaling and the statistical distribution of slopes. The experimental
range of anomalous scaling is studied in depth, and we investigate the effects of average
interface velocity, gap spacing, and disorder configuration. We also consider the situation in
which the interface is solely driven by capillary forces, which allows understanding better
the behavior of the interface at the scale of the disorder and its relation with the scaling
exponents in particular limits of the experiments.

The outline of the paper is as follows. In Sec.\ \ref{Sec:ANSATZS} we review the Family-Vicsek
and the anomalous scaling ansatzs. In Sec.\ \ref{Sec:setup} we introduce the experimental
setup, and in Sec.\ \ref{Sec:Data-Treatment} we describe the methodology used in data
analysis. The main experimental results are presented and analyzed in Sec.\ \ref{Sec:Results}.
The discussion and final conclusions are given in Section \ref{Sec:Conclusions}.

\section{Scaling ansatzs}\label{Sec:ANSATZS}

\subsection{Family-Vicsek scaling}

The statistical properties of a one--dimensional interface defined by a function $h(x,t)$
(interface height at position $x$ and time $t$) are usually described in terms of the
fluctuations of $h$. The {\it global} width in a system of lateral size $L$ is defined as
$w(L,t) = \langle [h(x,t) - \tilde{h}(t)]^2 \rangle_x ^{1/2}$, where $x$ is the position,
$\tilde{h}(t) = \langle h(x,t) \rangle_x$, and $< \; >_x$ is an spatial average in the $x$
direction. In the dynamic scaling assumption of Family-Vicsek (FV) \cite{Family-Vicsek-1985},
$w(L,t)$ scales as
\begin{equation}\label{global-width}
w(L,t)= L^{\alpha}f(L/t^{1/z}),
\end{equation}
where $z$ is the dynamical exponent and $f(u)$ the scaling function
\begin{equation}\label{FV-scaling}
f(u) \sim \left\{
\begin{array}{l}
\mbox{const} \qquad \mbox{if} \qquad u \ll 1, \\ u^{-\alpha} \qquad \mbox{if} \qquad u
\gg 1.
\end{array}
\right.
\end{equation}
Here $\alpha$ is the roughness exponent and characterizes the scaling of the surface at
saturation, $w(L,t \gg L^z) \sim L^{\alpha}$. $L^z$ characterizes the saturation time,
$t_{\times} \sim L^z$. It is customary to introduce a growth exponent $\beta$, which
characterizes the growth of the interfacial fluctuations before saturation, $w(L,t \ll L^{z})
\sim t^{\beta}$. The exponents $\alpha$, $\beta$, and $z$ verify the scaling relation
$\alpha=z \beta$.

An important feature of the FV assumption is that the scaling behavior of the surface can
also be obtained by looking at the {\it local} width over a window of size $l \ll L$. The
scaling is then
\begin{equation}\label{local-width}
w(l,t) \sim \left\{
\begin{array}{l}
t^{\beta} \qquad \mbox{if} \qquad t \ll l^z, \\ l^{\alpha} \qquad \mbox{if} \qquad t \gg
l^z.
\end{array}
\right.
\end{equation}
The roughness exponent $\alpha$ can also be obtained by studying the power spectrum of the
interfacial fluctuations, $S(k,t)=\langle H(k,t)H(-k,t)\rangle$, where $H(k,t)=\sum_x [h(x,t)
- \tilde{h}(t)] \exp (ikx)$. In the FV assumption, the power spectrum scales as
\begin{equation}\label{FV-power}
S(k,t)=k^{-(2\alpha+1)}s(kt^{1/z}),
\end{equation}
where $s$ is the scaling function
\begin{equation}\label{power-scaling}
s(u) \sim \left\{
\begin{array}{l}
u^{2\alpha+1} \qquad \mbox{if} \qquad u \ll 1, \\ \mbox{const} \qquad \mbox{if} \qquad u
\gg 1.
\end{array}
\right.
\end{equation}

\subsection{Anomalous scaling}

As mentioned in the introduction, experimental results and several growth models that do not
follow the FV dynamic scaling have led to the introduction of a new, more general dynamic
scaling hypothesis known as {\it anomalous scaling} \cite{Lopez-96,Lopez-97-I,Lopez-97-II}.
The origin of this new scaling comes from the different behavior of global and local
interfacial fluctuations. While the global width scales as in the FV scaling, the local width
follows a different scaling given by
\begin{equation}\label{Anomalous-width}
w(l,t) \sim \left\{
\begin{array}{l}
t^{\beta} \qquad \qquad \qquad \mbox{if} \qquad t \ll l^z, \\
l^{\alpha_{loc}}t^{\beta^{\ast}} \qquad \qquad \mbox{if} \qquad l^z \ll t \ll L^z,
\\ l^{\alpha_{loc}}L^{\alpha-\alpha_{loc}} \qquad \mbox{if} \qquad L^z \ll t,
\end{array}
\right.
\end{equation}
where $\beta^{\ast}=(\alpha-\alpha_{loc})/z$ is the local, anomalous growth exponent, and
$\alpha_{loc}$ the local roughness exponent. It is common to define
$\theta=\alpha-\alpha_{loc}$ as the {\it anomalous exponent}, which measures the degree of
anomaly \cite{Lopez-97-I,Lopez-97-II}. Therefore, in the case of anomalous scaling, two more
scaling exponents $\beta^{\ast}$ and $\alpha_{loc}$ are needed to characterize the scaling
behaviour of the surface. When $\theta=0$ and $\beta^{\ast}=0$ local and global scales behave
in the same way, and we recover the usual FV scaling.

It is common to define a scaling function for the behavior of $w(l,t)$, defining $g(u)$
as $g(l/t^{1/z})=w(l,t)/l^{\alpha}$. From Eq.\ (\ref{Anomalous-width}), $g(u)$ is
expected to scale as
\begin{equation}\label{Other-scaling-width}
g(u) \sim \left\{
\begin{array}{l}
u^{-(\alpha-\alpha_{loc})} \qquad \mbox{if} \qquad u \ll 1, \\ u^{-\alpha} \qquad \qquad
\mbox{if} \qquad u \gg 1.
\end{array}
\right.
\end{equation}

One of the implications of the anomalous scaling is that the local width saturates when the
system size saturates as well, i.e. at times $t_{\times}$, and {\it not} at the local time
$t_{l} \sim l^z$ as occurs in the FV scaling. It is also interesting to notice that the
anomalous scaling can be identified in an experiment by plotting the local width $w(l,t)$ as a
function of time for different window sizes. According to Eq.\ (\ref{Anomalous-width}) the
different curves would show a vertical shift proportional to $l^{\alpha_{loc}}$.

In terms of the power spectrum, Eq.\ (\ref{FV-power}), the anomalous scaling follows a new
scaling function $s(u)$ given by
\begin{equation}\label{anomalous-power-scaling}
s(u) \sim \left\{
\begin{array}{l}
u^{2 \alpha +1} \qquad \mbox{if} \qquad u \ll 1, \\ u^{2\theta} \qquad \qquad \mbox{if}
\qquad u \gg 1.
\end{array}
\right.
\end{equation}
For times $t \gg t_{\times}$ the power spectrum scales as $S(k,t) \sim
k^{-(2\alpha_{loc}+1)}t^{2\theta / z}$ and not simply as $k^{-(2\alpha+1)}$ of the
ordinary scaling.

In general, it is usual to differentiate between two types of anomalous scaling: {\it
super-roughness} and {\it intrinsic anomalous scaling}. Super-roughness is characterized by
$\alpha_{loc}=1$ and $\alpha > 1$, although the structure factor follows the ordinary scaling
given by Eqs.\ (\ref{FV-power}) and (\ref{power-scaling}). Thus, for super-roughness, the
power spectrum provides information about the global roughness exponent $\alpha$. The
intrinsic anomalous scaling follows strictly the scaling given by the scaling function of Eq.\
(\ref{anomalous-power-scaling}). This implies that the decay of the power spectrum with $k$
gives only a measure of $\alpha_{loc}$, not of $\alpha$. Moreover, the particular temporal
dependence of the scaling of the power spectrum allows identifying the existence of intrinsic
anomalous scaling by the presence of a vertical shift of the spectral curves at different
times. One has to be cautious with this particularity, because the observation of a vertical
shift, however, does not imply necessarily the existence of intrinsic anomalous scaling.

The different forms of anomalous scaling are included in a generic dynamic scaling ansatz
recently introduced by Ramasco {\it et al.} \cite{Ramasco-00}.

\begin{figure}
\begin{center}
\includegraphics[width=8.6cm]{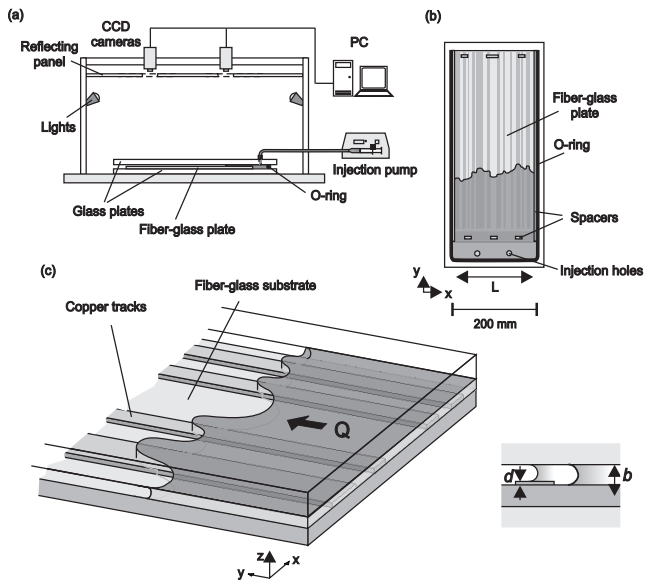} \vspace{0.5cm}
\caption{Sketch of the experimental setup. a) Side view. b) Top view. c) Schematic
representation of the oil--air interface and the disorder pattern.} \label{Fig:setup}
\end{center}
\end{figure}

\section{Experimental setup and procedure}\label{Sec:setup}

\begin{figure}
\begin{center}
\includegraphics[width=8.6cm]{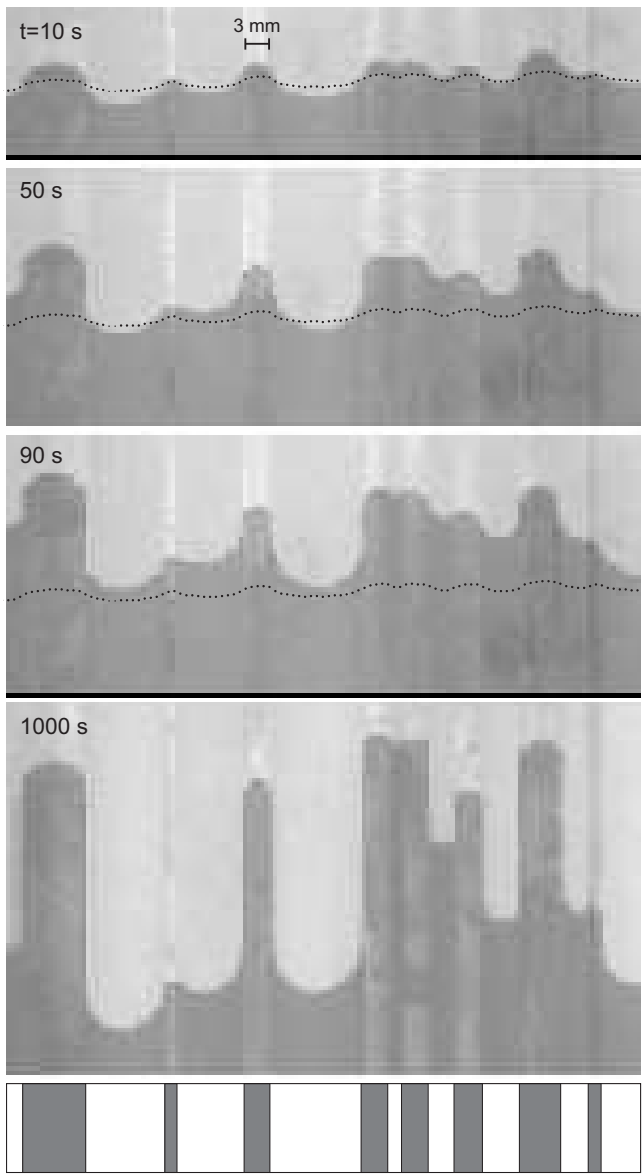} \vspace{0.5cm}
\caption{Close-up views of the oil--air interface with disorder T 1.50. The first three
pictures are early stages, while the last one corresponds to saturation. The position of the
copper tracks is shown in the bottom box. The dotted line shows the position and shape of the
interface profile at $t=0$ s. The experimental parameters are $v=0.08$ mm/s, $b=0.36$ mm.}
\label{Fig:close-up}
\end{center}
\end{figure}

The experimental setup used here has been described previously in Refs.\
\cite{Soriano-2002-I,Soriano-2002-PRL}. The setup (Fig.\ \ref{Fig:setup}) consists on a
horizontal Hele--Shaw cell which contains a pattern of copper tracks placed over a fiberglass
substrate. The tracks are d=$0.06 \pm 0.01$ mm high and are randomly distributed along the
lateral direction $x$ without overlap, and filling 35 \% of the substrate. The lateral size of
the unit track is $1.50 \pm 0.04$ mm, but wider tracks are obtained during design when two or
more unit tracks are placed one adjacent to the other. The separation between the fiberglass
substrate and the top glass defines the gap spacing $b$, which has been varied in the range
$0.16 \leq b \leq 0.75$ mm. A silicone oil (Rhodorsil 47 V) with kinematic viscosity $\nu=50$
mm$^2$/s, density $\rho=998$ kg/m$^3$, and surface tension oil--air $\sigma=20.7$ mN/m at room
temperature, is injected at one side of the cell, and displaces the air initially present. The
evolution of the oil--air interface is monitored using 2 CCD cameras. A typical example of the
morphology of an interface and its temporal evolution is shown in Fig.\ \ref{Fig:close-up}.

The oil is injected at constant volumetric injection rate $Q$ by means of a syringe pump. The
presence of the disorder, however, makes the cell inhomogeneous in the $z$ direction, so that
$Q$ is not exactly proportional to the average interface velocity $v$ measured on the images.
Nevertheless, since the free gap $b-d$ is always comparable to the total gap $b$, the {\it
two--dimensional} interface velocity $v$ is still well characterized. On the other hand, to
make the comparison between different experiments easy, we will measure $v$ in terms of a
reference velocity $V=0.04$ mm/s.

Although the disorder made of tracks (which will be called T 1.50, the numeric value
indicating the width of the unit track in mm) has been the main disorder pattern used in the
experiments, other disorder configurations have been used. These configurations are
characterized by a random distribution of squares with a lateral size of $1.50$ or $0.40$ mm,
and will be called SQ 1.50 and SQ 0.40 respectively. A full description and characterization
of these disorders, plus more detailed information about the experimental setup and procedure
can be found in Ref.\ \cite{Soriano-2002-I}.

\section{Data treatment}\label{Sec:Data-Treatment}

\begin{figure}
\begin{center}
\includegraphics[width=8.6cm]{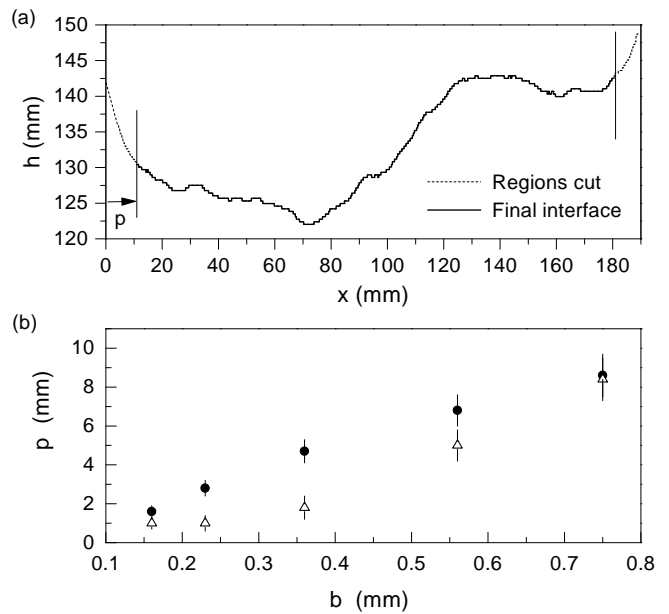} \vspace{0.5cm}
\caption{(a) Example of an interface with strong side wall distortions. The dashed lines
represent the regions cut at the end of the interface, and the solid line the interface
finally studied. The experimental parameters are $v=0.08$ mm/s, $b=0.75$ mm and disorder SQ
1.50. (b) Penetration length $p$ as a function of the gap spacing $b$ for SQ 1.50 (dots) and T
1.50 (triangles).} \label{Fig:p-b}
\end{center}
\end{figure}

As explained in Sec. \ref{Sec:ANSATZS}, we study the roughening process of the interfaces
through the interfacial {\it rms} width $w(l,t)$ and the power spectrum $S(k,t)$. The profile
$h(x,t)$, obtained from the analysis of the images, contains 515 points corresponding to a
lateral size of 190 mm. Because the oil tends to advance at the side walls of the cell, we
have a distorsion of the front that penetrates a distance $p$ from each side wall. This
penetration length depends on average interface velocity and is specially sensitive to
variations in gap spacing. An example is presented in Fig.\ \ref{Fig:p-b}(a), which shows the
effect of the distortion on an interface with experimental parameters $v=0.08$ mm/s, $b=0.75$
mm and disorder SQ 1.50. To evaluate the effect of the distortion, we have measured the
penetration length $p$ at different gap spacings $b$ for two disorder configurations, SQ 1.50
and T 1.50. The results are presented in Fig.\ \ref{Fig:p-b}(b), and show that $p$ increases
quickly with gap spacing. For $b=0.75$ mm, the largest value studied, $p$ is about 8 mm.
Notice also that the presence of a columnar disorder reduces significantly the effect of the
distortion for the smallest gap spacings. To minimize the effect of the distortion we have cut
8 mm from both ends of all interfaces, as shown in Fig.\ \ref{Fig:p-b}(a), reducing them to a
final lateral size of 174 mm (470 pixels). For data analysis convenience (particularly in the
computation of the power spectrum), we have extended the number of points to 512 using linear
interpolation. Finally, we have forced periodic boundary conditions by subtracting the
straight line that connects the two ends of the interface. This procedure, well documented in
the literature of kinetic roughening \cite{Simonsen98}, eliminates an artificial overall slope
-2 in the power spectrum due to the discontinuity at the two ends \cite{Schmittbuhl95}.

Because the initial stages of roughening are so fast, the interface profile at $t=0$ s is not
a flat interface, but a profile with an initial roughness $w(0)$, as can be observed in Fig.\
\ref{Fig:close-up}. This roughness, on one hand, does not permit to observe the growth of the
small scales at short times; on the other, $w(0)$ is different from one disorder configuration
to another, which makes the early stages strongly dependent on the particular realization of
the disorder. To eliminate these effects, we will always plot the {\it subtracted width}
$W(l,t)$, defined as $W(l,t)=\langle w^2(l,t) - w^2(l,0) \rangle ^{1/2}$, where the brackets
indicate average over disorder realizations. This correction is an standard procedure in the
field of growing interfaces, and is reported in both experiments and numerical simulations
\cite{Barabasi-Stanley,Sputtering-Jeffries,Tripathy00}.

Another aspect of interest concerning the analysis of our experimental data is the various
methods available in the literature to obtain the growth and roughness exponents. In this work
we have obtained the growth exponents $\beta$ and $\beta^{\ast}$ through the scaling of
$W(l,t)$ with time. We have also used an alternative way to measure $\beta^{\ast}$, proposed
by L\'opez \cite{Lopez-slopes-99}, in which the growth of the mean local slopes, $\rho (t) =
\langle \overline{(\nabla h)^2} \rangle^{1/2}$, is analyzed. Usually $\rho (t)$ scales with
time as $\rho (t)\sim t^{\kappa}$, with $\kappa < 0$ for the ordinary scaling and $\kappa > 0$
for anomalous scaling. In this last case, $\kappa$ is identified with the local growth
exponent $\beta^{\ast}$. The local roughness exponents $\alpha_{loc}$ have been obtained
through the analysis of the power spectrum $S(k,t)$. The results obtained have been checked
using other methods, such as the dependency of the height-height correlation function $C(l,t)$
or $W(l,t)$ with the {\it window} size $l$, in both cases scaling as $l^{\alpha_{loc}}$. We
have obtained, within error bars, similar results to those from $S(k,t)$. Finally, in the
framework of intrinsic anomalous scaling, the only direct way to measure the global roughness
exponents $\alpha$, and in addition $z=\alpha / \beta$, is through the analysis of $C(L,t)$ or
$W(L,t)$ with the {\it system} size $L$. Both quantities scale as $L^{\alpha}$. Because it is
not practical to perform experiments varying the system size, we have not measured global
roughness exponents directly. Instead, they have been obtained from the collapse of $S(k,t)$
or $W(l,t)$. Notice that information about the value of $\alpha$ is contained in both the
vertical displacement of the spectral curves with time (Eqs.\ (\ref{FV-power}) and
(\ref{anomalous-power-scaling})) and the vertical shift of the $W(l,t)$ curves for different
window sizes (Eq.\ (\ref{Anomalous-width})). The error bars that we give in the final numeric
value of an specific scaling exponent takes into account the various methods of analysis and
the error in the different collapses.

\section{Experimental results and analysis}\label{Sec:Results}

\subsection{Evidence of anomalous scaling}

\begin{figure}
\begin{center}
\includegraphics[width=8.6cm]{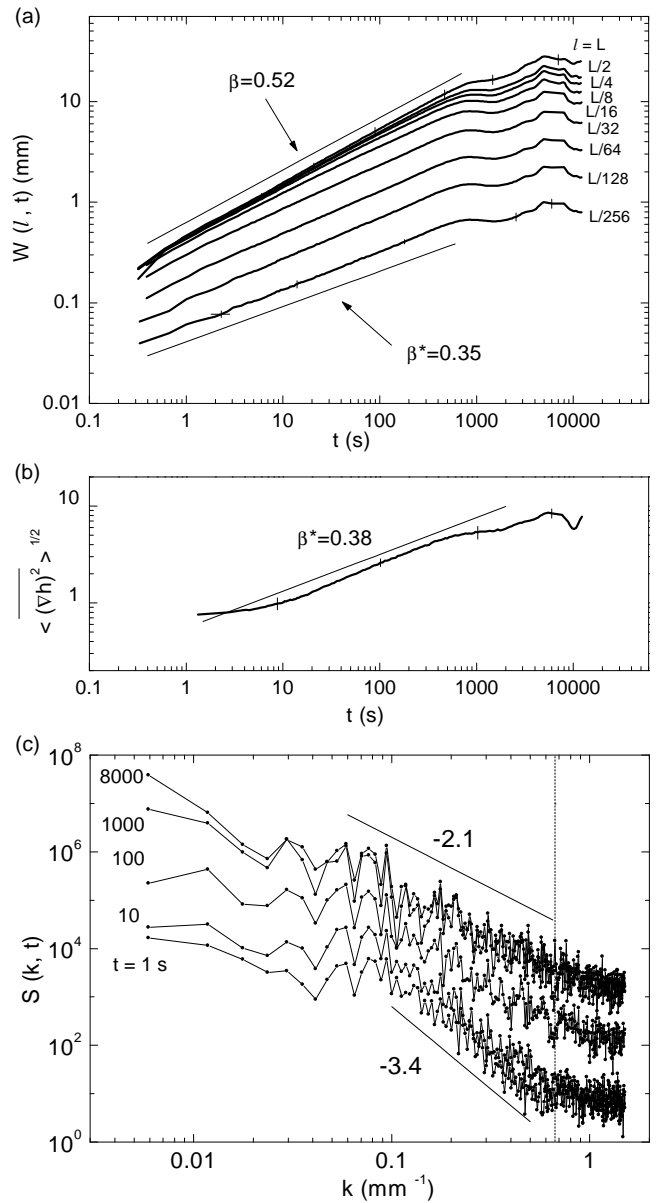} \vspace{0.5cm}
\caption{Experimental results for $b=0.36$ mm, $v=0.04$ mm/s ($V$) and disorder T 1.50.
a) Analysis of $W(l,t)$ for different window sizes. b) Analysis of the mean local slopes
as a function of time. c) Temporal evolution of the power spectrum. The vertical line in
c) indicates the spatial scale of the disorder.} \label{Fig:T-V}
\end{center}
\end{figure}

The first evidence of anomalous scaling in these experiments was presented in our previous
work \cite{Soriano-2002-PRL} and corresponded to the experimental parameters $v=0.08$ mm/s
($2V$) and $b=0.36$ mm. Here, for completeness, we will show similar experimental results for
a different set of parameters, namely $v=0.04$ mm/s ($V$) and $b=0.36$ mm.

\subsubsection{Interfacial width}

Fig.\ \ref{Fig:T-V}(a) shows the results obtained for
$W(l,t)$ for progressively smaller window sizes. We obtain clear exponents $\beta=0.52
\pm 0.04$ (corresponding to $l=L$) and $\beta^{\ast}=0.35 \pm 0.05$ (for $l=L/256$). The
fact that $\beta^{\ast}$ is clearly different from zero is a robust evidence of anomalous
scaling. Moreover, the behavior of the different curves is in good agreement with the
anomalous scaling ansatz of Eq.\ (\ref{Anomalous-width}), i.e. the plots for different
window sizes are vertically shifted. Notice also that the saturation takes place at
$t_{\times}$ (the saturation time of the full system) for all window sizes.

\subsubsection{Local slopes}

To check the reliability of the exponent $\beta^{\ast}$ we have also studied the scaling of
the mean local slopes with time. The result is presented in Fig.\ \ref{Fig:T-V}(b), obtaining
$\beta^{\ast}=0.38 \pm 0.05$, in agreement with the result of $\beta^{\ast}$ using
$W(l=L/256,t)$.

\subsubsection{Power spectrum}

The analysis of the interfacial fluctuations through the power spectrum is presented in Fig.\
\ref{Fig:T-V}(c). The first important observation is that the spectral curves are shifted
vertically, the shift being maximum at short times, reducing progressively as time increases,
and disappearing at saturation ($t = t_{\times} \simeq 2000$ s). The presence of this vertical
displacement of the spectral curves is a signature of anomalous scaling. The evolution of the
power spectrum gives two clear power law dependencies, at short times with slope $-3.4 \pm
0.1$, and at saturation with slope $-2.1 \pm 0.1$. The former is a transient regime whose
origin is the strong dominance of the capillary forces at very short times. The latter
provides the local roughness exponent, $\alpha_{loc} = 0.55 \pm 0.10$ in this case.

\begin{figure}
\begin{center}
\includegraphics[width=8.6cm]{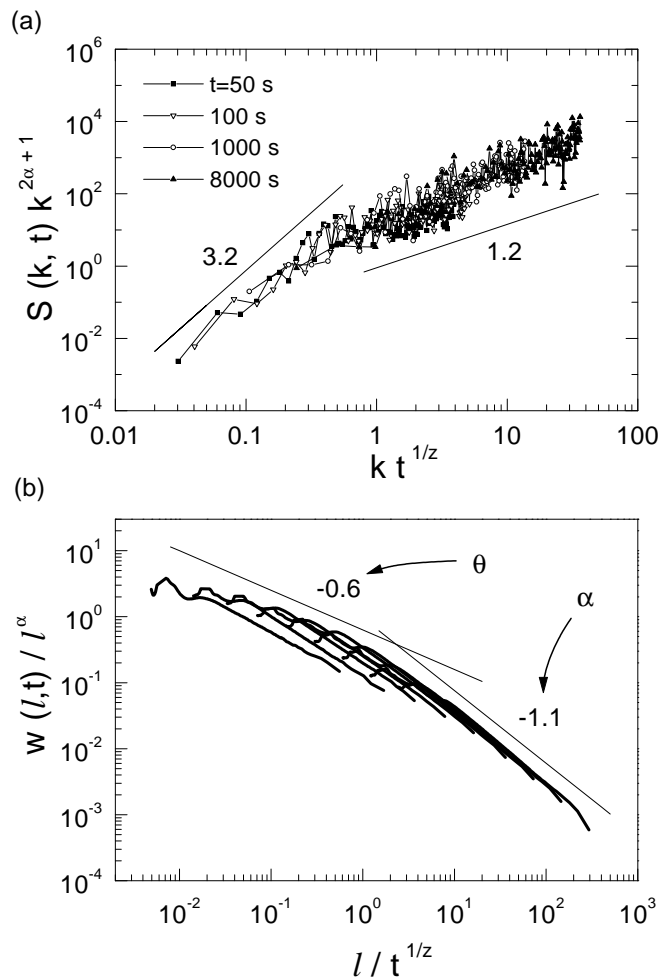} \vspace{0.5cm}
\caption{a) Collapse of the power spectrum of Fig.\ \ref{Fig:T-V}(c) for $t>10$ s using the
scaling function of Eq.\ (\ref{anomalous-power-scaling}). b) Collapse of the $W(l,t)$ plots of
Fig.\ \ref{Fig:T-V}(a), using the scaling function of Eq.\ (\ref{Other-scaling-width}).}
\label{Fig:collapse}
\end{center}
\end{figure}

To get the values of $\alpha$ and $z$ we have collapsed the plots of Fig.\ \ref{Fig:T-V}(c) in
accordance with the scaling function of Eq.\ (\ref{anomalous-power-scaling}), i.e. plotting
$S(k,t)k^{2 \alpha +1}$ as a function of $k t^{1/z}$. This collapse, shown in Fig.\
\ref{Fig:collapse}(a), refines the numerical values of the experimental results and gives the
values of $\alpha$ and $z$. We obtain in this way:
\begin{eqnarray}\label{results}
\nonumber & & \beta=0.50 \pm 0.04,\; \beta^{\ast}=0.30 \pm 0.04,\\
          & & \alpha=1.1 \pm 0.1,\;\alpha_{loc}=0.5 \pm 0.1,\; z=2.2 \pm 0.2.
\end{eqnarray}

An alternative collapse that confirms the validity of these results can be performed
using the scaling function defined in Eq.\ (\ref{Other-scaling-width}). This collapse is
shown in Fig.\ \ref{Fig:collapse}(b).

\subsubsection{Multiscaling}

\begin{figure}
\begin{center}
\includegraphics[width=8.6cm]{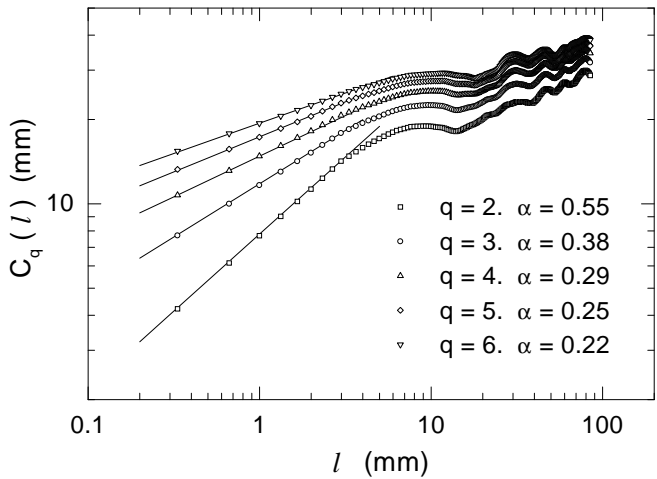} \vspace{0.5cm}
\caption{Multiscaling analysis through the computation of the q--order height-height
correlation function for interfaces at saturation. The experimental parameters are $v=0.04$
mm/s, $b=0.36$ mm and disorder T 1.50.} \label{Fig:multiscaling}
\end{center}
\end{figure}

The presence of anomalous scaling can also be identified by checking for multiscaling
\cite{krug}. For this purpose, we have studied the scaling of the generalized correlations of
order $q$, of the form $C_q (l) = \lbrace {(h(x+l,t)-h(x,t))^{q}}\rbrace^{1/q} \sim
l^{\alpha_{q}}$, where $\alpha_q$ are local roughness exponents. The results obtained are
shown in Fig.\ \ref{Fig:multiscaling}, which gives the values of $\alpha_{q}$ at different
orders $q$. We get a progressive decrease of the measured exponent, varying from $\alpha_{2}
\simeq 0.55$ to $\alpha_{6} \simeq 0.22$, which confirms the existence of multiscaling in the
experiment and, therefore, of anomalous scaling.

\subsection{Experimental range of anomalous scaling}

Once the existence of anomalous scaling is well characterized for a specific set of
experimental parameters, we are now interested in analyzing the range of validity of the
anomaly. We have performed a series of experiments varying $v$ and fixing the gap spacing to
$b=0.36$ mm and, on the other hand, a series of experiments varying $b$ and fixing the average
interface velocity to $v=0.08$ mm/s (2V). Because we can identify the existence of anomalous
scaling by looking at the behavior of the smallest scales, i.e. the existence of a
$\beta^{\ast} \neq 0$, we present the results by plotting the local growth exponent
$\beta^{\ast}$ for the different experiments.

\subsubsection{Interface velocity}

\begin{figure}
\begin{center}
\includegraphics[width=8.6cm]{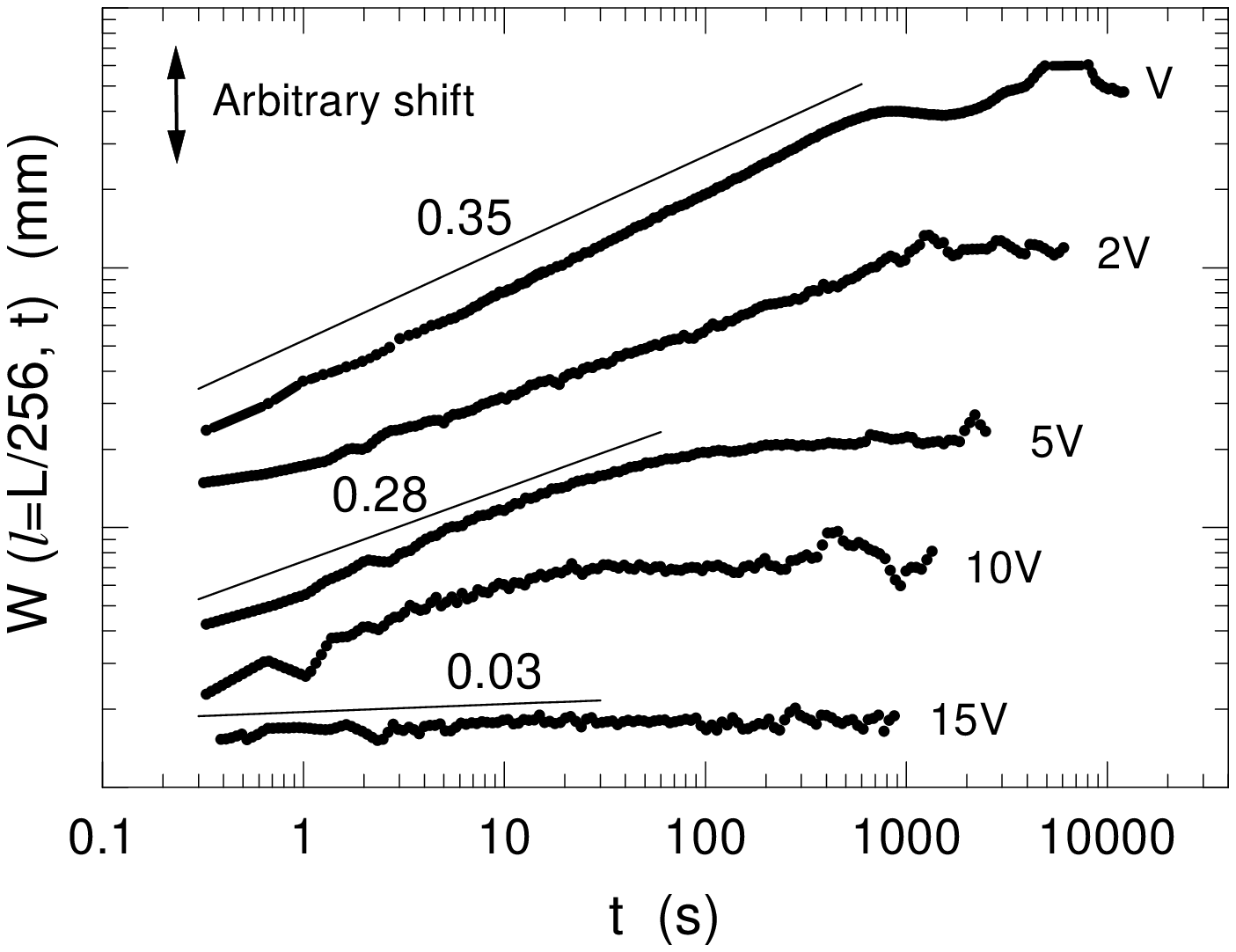} \vspace{0.5cm}
\caption{W(t) curves for a window size $l=L/256$ and different interface velocities. The gap
spacing is $b=0.36$ mm in all cases, and $V=0.04$ mm/s. The curves are vertically shifted for
clarity.} \label{Fig:local-v}
\end{center}
\end{figure}

The results of $\beta^{\ast}$ at different $v$ are presented in Fig.\ \ref{Fig:local-v}.
In a wide range of interface velocities, from $V$ to $10\,V$, the measured exponent
$\beta^{\ast}$ is robust and not very sensitive to changes in the velocity, varying from
$0.35 \pm 0.04$ for velocity $V$ to $0.28 \pm 0.04$ for velocity $10\,V$. For velocities $v
\gtrsim 15V$, however, the measured value of $\beta^{\ast}$ quickly decays to a value
close to zero, indicating that over a threshold velocity (about $13V$ for $b=0.36$ mm)
the anomaly disappears.

In terms of the power spectrum, when the velocity is varied from $V$ to $15V$, we always
observe a vertical shift of the spectral curves, and the power spectrum at different times can
be collapsed to get the set of scaling exponents. Above the previous threshold velocity, the
power spectrum follows the ordinary scaling, i.e. without vertical displacement.

\subsubsection{Gap spacing}

\begin{figure}
\begin{center}
\includegraphics[width=8.6cm]{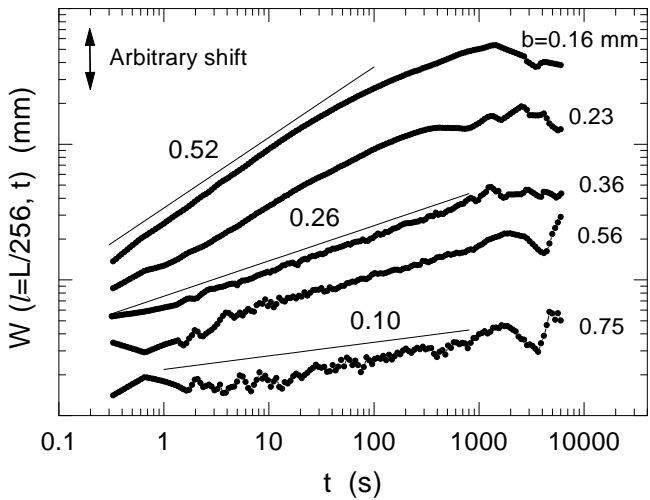} \vspace{0.5cm}
\caption{W(t) curves for a window size $l=L/256$ and different gap spacings. The interface
velocity is $v=2V=0.08$ mm/s in all cases. The curves are vertically shifted for clarity.}
\label{Fig:local-b}
\end{center}
\end{figure}

The results of $\beta^{\ast}$ at different $b$ are presented in Fig.\ \ref{Fig:local-b}.
It is clear from the different plots that the exponent $\beta^{\ast}$ is rather sensitive
to variations of the gap spacing. Three different behaviors can be observed: i) For {\it
small} values of $b$ (around $b=0.16$ mm), where capillary forces are dominant, we get
$\beta^{\ast} = 0.52 \pm 0.04$, which is identical to the corresponding value of $\beta$.
This similitude is discussed in Sec.\ \ref{Subsec:Regular}. ii) For {\it moderate} values
of $b$ (in the range $0.23 \leq b \leq 0.56$ mm) we get values of $\beta^{\ast}$ that
vary from $0.35 \pm 0.05$ for $b=0.23$ mm to $0.25 \pm 0.04$ for $b=0.56$ mm. This range
of gap spacings is characterized by strong (but not dominant) capillary forces, and is
the region where the anomalous scaling can be fully characterized. iii) For {\it large}
values of $b$ ($b \gtrsim 0.60$ mm) we get $\beta^{\ast} = 0.10 \pm 0.04$ for $b=0.75$
mm, and $\beta^{\ast}$ is nearly zero for larger gap spacings. In this regime capillary
forces are not sufficiently strong, and the anomalous scaling disappears. The analysis of
$\beta^{\ast}$ through the scaling of the mean local slopes has given identical results
within error bars, although the sensitivity of this analysis to short--time transients
makes it less reliable.

Concerning the power spectrum, in the range $0.23 \leq b \leq 0.56$, the vertical shift
of the spectral curves is present, and a regime characterized by anomalous scaling can be
described. For $b=0.16$ mm, although the spectral curves are vertically shifted, an
scaling regime is not identifiable at all. The power spectra can be collapsed, but not
reliable $\alpha$, $\alpha_{loc}$, and $z$ exponents can be obtained. For $b \gtrsim 0.6$
mm, a vertical shift in the power spectra is not observed, and the scaling of the
interfacial fluctuations is best described in the framework of ordinary scaling.

\subsubsection{Disorder properties}

\begin{figure}
\begin{center}
\includegraphics[width=8.6cm]{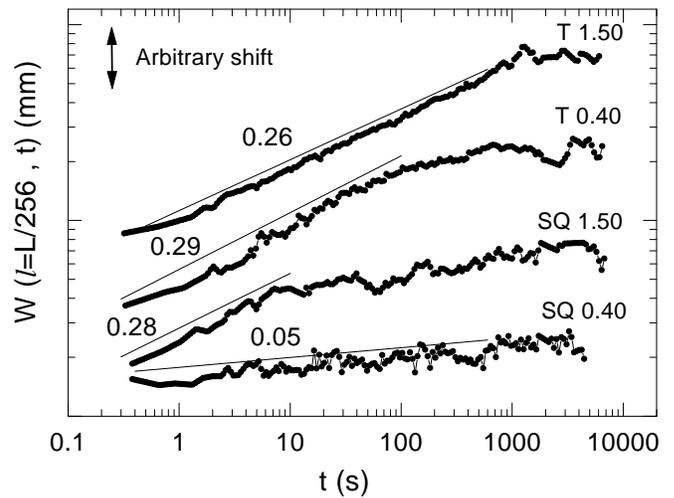} \vspace{0.5cm}
\caption{W(t) curves for a window size $l=L/256$ and different disorder configurations with
progressively shorter persistence in the growth direction. $b=0.36$ mm and $v=2V=0.08$ mm/s in
all cases. The curves are vertically shifted for clarity.} \label{Fig:T-to-SQ}
\end{center}
\end{figure}

Other modifications that we have carried out to study the range of validity of the anomalous
scaling are the track width and the persistence of the disorder along the growth direction.
Fig.\ \ref{Fig:T-to-SQ} shows the variation of the local interfacial fluctuations for four
different disorder configurations. The gap spacing and velocity have been fixed to $b=0.36$ mm
and $v=2V$ respectively. For T 0.40 we get a similar result from that obtained with T 1.50,
measuring $\beta^{\ast}=0.29 \pm 0.05$. However, we have evidences that the anomaly decreases
as we reduce the width of the basic track, and it disappears when the width of the basic track
is $\lesssim 0.20$ mm, due to the fact that the in--plane curvature over the copper track
becomes similar to the curvature in the $z$ direction.

Next, we study different disorder configurations with progressively smaller persistence in the
growth direction. For SQ 1.50, although the saturation time for the full system is at
$t_{\times} \simeq 1000$ s, we observe that the local interfacial fluctuations saturate at
much shorter time, which is more characteristic of the ordinary scaling. However, we still
observe traces of anomalous scaling at early times, for which $\beta^{\ast}=0.26 \pm 0.04$.
This is somehow not surprising taking into account that, for this disorder, the first array of
squares is equivalent to T 1.50, but only $1.50$ mm long. Thus, at very short times, the
growth follows the same behavior as tracks. With a driving velocity $0.08$ mm/s, the interface
takes about $10-20$ s in crossing the first array of squares, which is the same time for which
we observe the presence of anomalous scaling. For SQ 0.40, with $t_{\times} \simeq 300$ s, we
measure $\beta^{\ast}=0.05 \pm 0.04$, which is indicative of the absence of anomalous scaling.

\subsubsection{Characterization by the statistical distributions of slopes}

\begin{figure}
\begin{center}
\includegraphics[width=8.6cm]{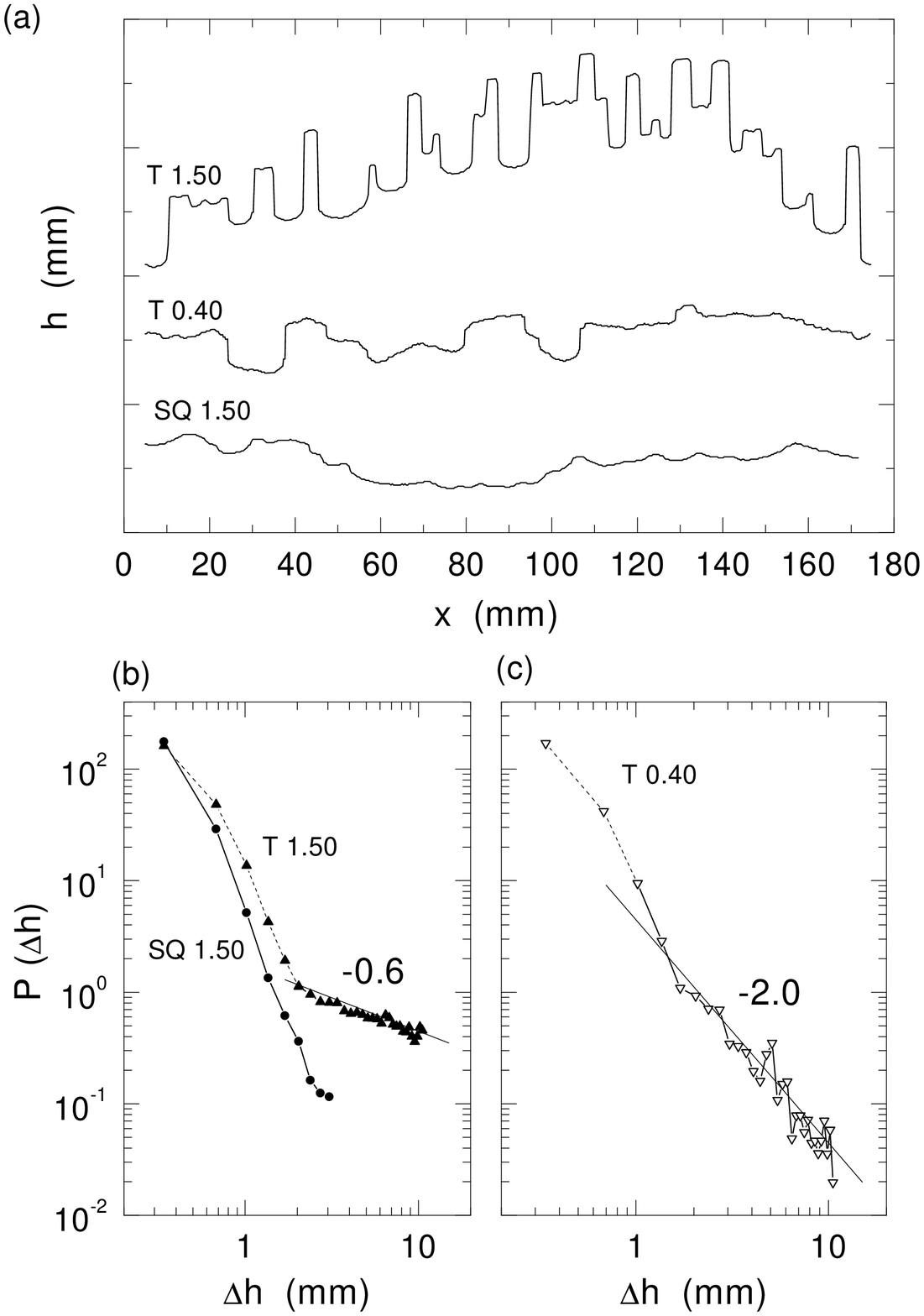} \vspace{0.5cm}
\caption{(a) Samples of interfaces for different disorder configurations and identical
experimental parameters $v=0.08$ mm/s and $b=0.36$ mm. (b) Distribution function $P(\Delta
h)$, with $\Delta h \equiv | h(x_{i+1} - h_x |$, for local slopes, comparing T 1.50 and SQ
1.50. (c) The same for T 0.40. See text for details of the power law fits.} \label{Fig:histo}
\end{center}
\end{figure}

We have also considered an alternative analysis to confirm the results of anomalous
scaling. In a recent work, Asikainen {\it et al.} \cite{Histogram-slopes-Asikainen}
showed that the presence of anomalous scaling can be associated to the presence of
multiscaling on one hand, and to high slopes in the morphology of the interfaces on the
other. According to these authors, the distribution function of these slopes, $P(\Delta
h)$, with $\Delta h \equiv | h(x_{i+1} - h_x |$, scales following a L\'evy distribution,
$P(\Delta h) \sim \Delta h^{- \gamma}$, where $0 < \gamma < 2$; while for ordinary
scaling $P(\Delta h)$ follows a Gaussian distribution. We have checked this prediction
with our experimental results. Fig.\ \ref{Fig:histo}(a) compares the shape of three
different interfaces with different degrees of anomalous scaling. In all cases $v=2V$ and
$b=0.36$ mm. All the interfaces are in the saturation regime. For T 1.50 (the case most
representative of anomalous scaling) the interface presents high slopes at the edges of
the copper tracks. For T 0.40 some regions present high slopes, but in general the
interface looks smoother. This case shows anomalous scaling, but the degree of the
anomaly is much weaker than the previous case. For SQ 1.50 the interface is very smooth
and the interfacial fluctuations are best described with the FV scaling, with only traces
of anomalous scaling at very short times. The interfaces in these three cases are similar
to those reported in fracture of wood by Eng{\o}y {\it et al.} in Fig. 1 of Ref.\
\cite{fracture-Wood-Engoy}. In particular, the interfaces for T 1.50 or T 0.40 look like
interfaces obtained from tangential fracture in wood, both characterized by high slopes
and anomalous scaling. In the same line, the interfaces for SQ 1.50 are similar to those
obtained from radial fracture. These similitudes indicate that the role of the wood
fibers is probably equivalent to the role of the copper tracks in our system as the
origin of the anomalous scaling.

The distribution functions for the different interface morphologies are shown in Fig.\
\ref{Fig:histo}(b) and (c). For T 1.50 we have two different regimes. Up to $\Delta h \simeq
2$ mm (dashed line in the plot) we have the contribution of the slopes corresponding to the
structure inside tracks. For $\Delta h > 2$ mm (solid line), the contribution to the
distribution function arises from the edges of the tracks. This last contribution is the one
responsible for the anomalous scaling, and scales with $\Delta h$ with an exponent $-0.6 \pm
0.1$. A similar result is observed for T 0.40. In this case, if we do not consider the
contribution of the structure inside tracks (up to $1$ mm), we get a power law dependency with
exponent $-2.0 \pm 0.2$. The fact that the anomaly is weaker for T 0.40 compared to T 1.50 is
illustrated in this case by the larger value of the slope. If we compare the distribution
functions of T 1.50 and SQ 1.50 we observe that, in the latter case, there is contribution
only from small slopes. $P(\Delta h)$ would scale in this case with an exponent $-3.6$, far
from the values expect for anomalous scaling.

\subsection{Experiments at Q=0}

In order to understand the origin of the anomalous scaling, the previous results indicate that
a more detailed investigation about the role of the capillary forces at the scale of the
disorder has to be carried out. For this reason we have performed experiments at $Q=0$, where
interfacial growth is driven solely by capillary forces.

The experiments at $Q=0$ have been performed by initially driving the interface at high
velocity until it has reached the center of the cell. Then the pump is stopped and the
experiment started. We have investigated five gap spacings, in the range $0.16 \leq b \leq
0.75$ mm, and characterized the interface velocity over copper tracks and fiberglass channels.
For sufficiently small gap spacing, up to $b \simeq 0.36$ mm, the morphology of the interface
adopts a finger--like configuration, which ultimately causes that the interface pinches--off
at long times. For this reason we have reduced the duration of the experiments to $500$ s for
all gap spacings.

\begin{figure}
\begin{center}
\includegraphics[width=8.6cm]{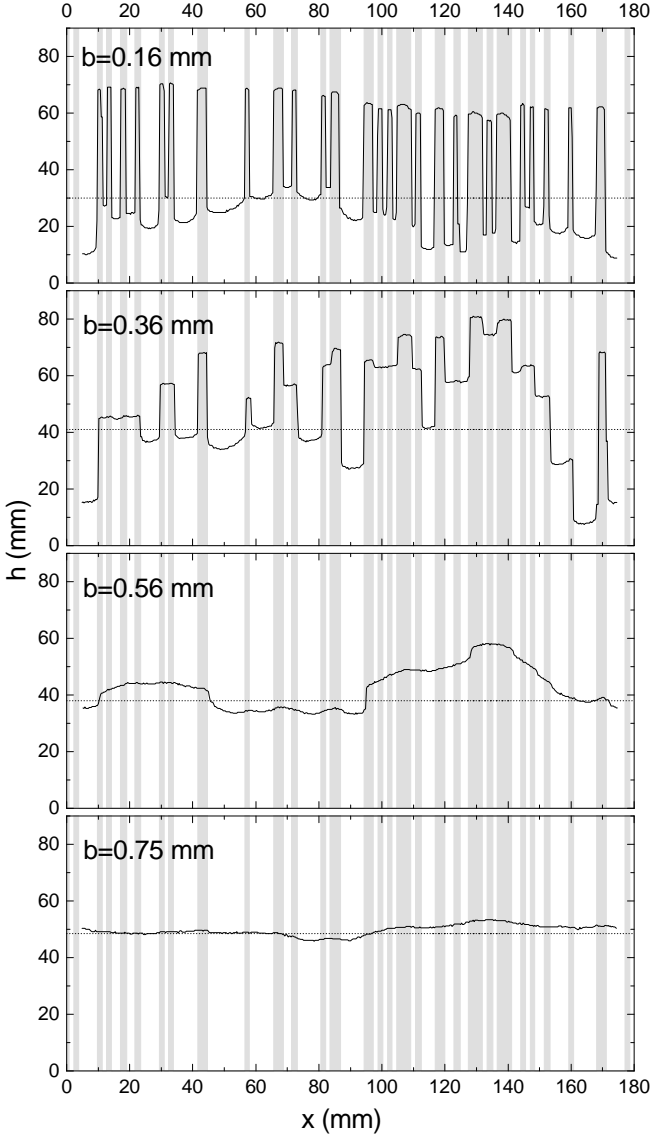} \vspace{0.5cm}
\caption{Example of interfaces at $Q=0$ with different gap spacings and disorder T 1.50
mm. All the interfaces correspond to $t=500$ s. The horizontal dotted line indicates the
position of the interface at $t=0$ s.} \label{Fig:q0}
\end{center}
\end{figure}

Fig.\ \ref{Fig:q0} shows a comparison of the interfaces at the end of the experiments with
different gap spacings. The horizontal line represents the average position of the interface
at $t=0$. For $b=0.16$ mm the points of the interface over the different tracks move almost
independently, with weak coupling between neighboring tracks: the oil over copper tracks
always advances, while the oil over fiberglass channels, except for some particular locations,
always recedes, driven by the different capillary forces associated with the different
curvature of the meniscii over copper tracks and fiberglass substrate. As we increase the gap
spacing, the coupling in the motion of the interface over neighboring tracks is progressively
more intense. For $b=0.36$ mm the advance over a specific copper track is coupled with its
neighbors. The oil advances or recedes in a given track or channel depending on the particular
realization of the disorder. Only the widest copper tracks or fiberglass channels have an
independent motion. For $b = 0.56$ mm the coupling reaches larger regions of the cell, and for
$b=0.75$ the capillary forces are not longer sufficient to allow a large deformation of the
interface.

It is also noticeable that the average position of the interface does not keep at rest.
Indeed, it advances in the same direction as the oil over copper tracks. This is a consequence
of the three-dimensional nature of the experiment. The total mass of fluid in the cell is
conserved, but not the area measured on the images. However, the measured velocity of advance
of the interface is very low. For $b=0.36$ mm, for example, the velocity is about $V/10$.

\begin{figure}
\begin{center}
\includegraphics[width=8.6cm]{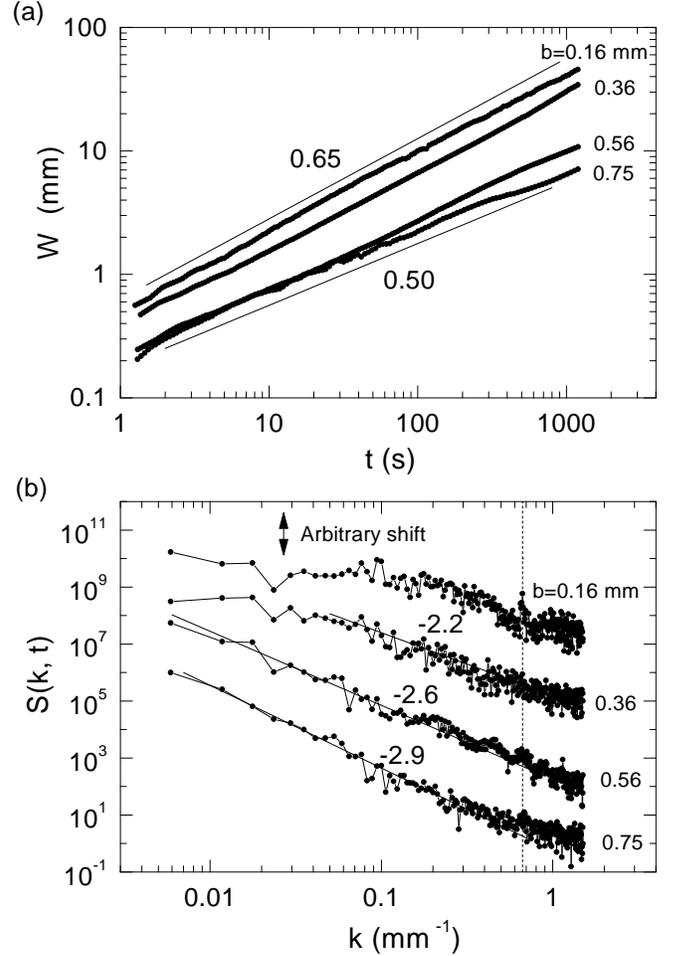} \vspace{0.5cm}
\caption{Log-log plot of $W(t)$ for experiments at $Q=0$ with four different gap spacings and
disorder T 1.50 mm. The straight line with slope $0.65$ corresponds to a fit of the
experiments with $b=0.16$ mm and $0.36$ mm. The straight line with slope $0.50$ is a fit to
the experiment with $b=0.75$ mm. Each curve is an average over two disorder realizations, with
one run per disorder realization.} \label{Fig:q0-w-t}
\end{center}
\end{figure}

The first analysis that we have performed with the experiments at $Q=0$ is studying the growth
of interfacial fluctuations with time. Fig.\ \ref{Fig:q0-w-t} shows the $W(t)$ curves for four
different gap spacings. We observe power laws with a growth exponent $\beta$ that decreases
from $0.65 \pm 0.04$ for $b=0.16$ mm to $0.50 \pm 0.04$ for $b=0.75$ mm. Clearly, when the
capillary forces are sufficiently strong to allow for both the advance of the oil over tracks
and the recession of the oil over fiberglass channels, we get a larger value of the growth
exponent than in all other cases. The same behavior is observed for $\beta^{\ast}$, varying
from $0.65 \pm 0.04$ for $b=0.16$ mm to $0.15 \pm 0.05$ for $b=0.75$ mm.

Concerning the behavior of the power spectrum, the fact that the experiments do not reach a
saturation regime makes the analysis of the power spectrum incomplete. Qualitatively, we have
observed a vertical shift of the spectral curves as time increases for gap spacings up to
$b=0.56$ mm. For $b=0.75$ mm the vertical shift is not present, confirming the absence of
anomalous scaling for values of $b \geq 0.60$ mm for any interface velocity.

\begin{figure}
\begin{center}
\includegraphics[width=8.6cm]{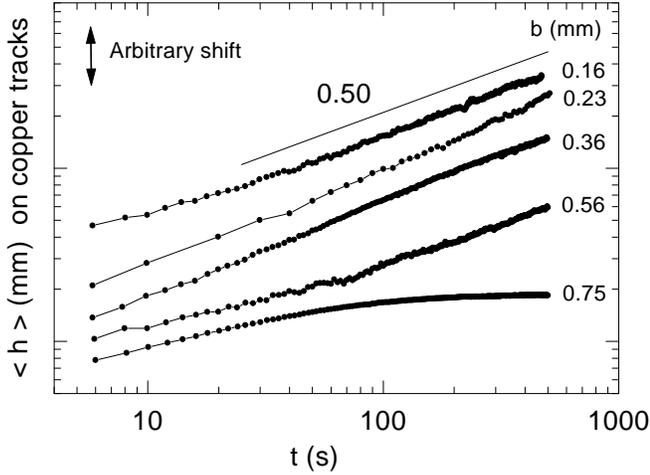} \vspace{0.5cm}
\caption{Log-log plot of the average height as a function of time for the interface points
advancing over copper tracks, for different gap spacings. $Q=0$ in all cases. The curves are
vertically shifted for clarity.} \label{Fig:heigth-copper}
\end{center}
\end{figure}

The second analysis that we have performed has been a detailed study of the motion of the
interface over copper tracks or fiberglass channels separately. In Fig.\
\ref{Fig:heigth-copper} we represent the variation of the mean height $<h>$ for points of
the interface that are advancing over copper tracks. For gap spacings in the range $0.16
\leq b \leq 0.36$ mm, the mean height is a power law of time with an exponent $0.50 \pm
0.03$. For larger gap spacings, the exponent is progressively smaller, and reaches $0.32
\pm 0.03$ for $b=0.75$ mm. We have hints that the exponent tends to zero as the capillary
forces become even weaker. A similar analysis can be carried out for the fiberglass
channels. In this case we get the same behavior in the range $0.16 \leq b \leq 0.36$ mm,
with an exponent $0.50 \pm 0.03$. For $b \geq 0.60$ mm not clear scaling can be
identified. Finally, we have also studied the motion on individual tracks, in order to
know the role of track width. Considering tracks whose motion is not coupled with its
neighbors we have obtained that tracks of different width show the same behavior $<h>
\sim t^{\, 0.5}$, but the different curves are displaced a factor proportional to the
track width.

From these studies we conclude that it is possible to describe the motion over a single copper
track, $v^i_+$, or fiberglass channel, $v^i_-$, by the simple law
\begin{equation}\label{local-velocity}
  v^i_{\pm} (t) = A^i_{\pm} (b, D^i_{\pm}) \; t^{-1/2} \qquad \mbox{for} \qquad t \geqslant 1 \mbox{ s},
\end{equation}
where $A^i_{\pm} (b,D^i_{\pm})$ is a prefactor that depends on the gap spacing $b$ and
the individual track or channel width $D^i$. Averaging out the prefactor $A^i_{\pm}$ over
different track or channel widths we obtain a characteristic maximum velocity $v_{M\pm}$
for a given gap spacing. Eq.\ (\ref{local-velocity}) becomes now
\begin{equation}\label{Eq:Washburn}
  v_{\pm} (t) = v_{M\pm} (b) \; t^{-1/2} \qquad \mbox{for} \qquad t \geqslant 1 \mbox{ s}.
\end{equation}
In general $v_{M_+} > v_{M_-}$ due to the different gap spacing on copper tracks and on
fiberglass channels. This difference, however, is progressively smaller as $b$ increases. For
$b \geq 0.36$ mm it is reasonable to consider $v_{M_+} \simeq v_{M_-} \equiv v_M$. The time
$t=1$ s, determined from systematic experimental observations,  is a typical time in which the
velocity jumps from the average interface velocity out of the disorder to $v_M$ inside the
disorder.

%% This maximum velocity is the result of the fast jump experienced by the local interfacial
%% velocity when the interface gets in contact with the disorder for the first time.

\subsection{From $Q=0$ to low injection rates}

Eq.\ (\ref{Eq:Washburn}) gives the velocity over copper tracks or fiberglass channels when
$Q=0$ (no injection). In the presence of injection, when the interface is driven at an average
interface velocity $v \neq 0$, the behavior observed experimentally is modified in the form
\begin{equation}\label{Eq:Washburn-v}
  v_{\pm} (t) = v \pm (v_M - v) \; t^{-1/2},
\end{equation}
where $v_M=v_M(b)$ and $v_+$ and $v_-$ are the velocities averaged over copper tracks and
fiberglass channels respectively. Thus, when the interface gets in contact with the disorder,
the velocity on copper tracks jumps to a maximum $v_M$ in a characteristic time of $1$ s and
then relaxes as $t^{-1/2}$ to the nominal velocity. On fiberglass channels, the velocity is
zero or even negative at short times, and as time goes on, the velocity increases until it
reaches the nominal velocity at long times. Fig.\ \ref{Fig:close-up} illustrates this
behavior. At short times, $t \simeq 10$ s, all the oil over copper tracks is advancing, while
the oil on the widest fiberglass channels is receding. At $t \simeq 50$ s the oil on
fiberglass channels has stopped receding. At later times, the velocity on different points of
the interface tends asymptotically to the nominal velocity, reached at saturation.

\begin{figure}
\begin{center}
\includegraphics[width=8.6cm]{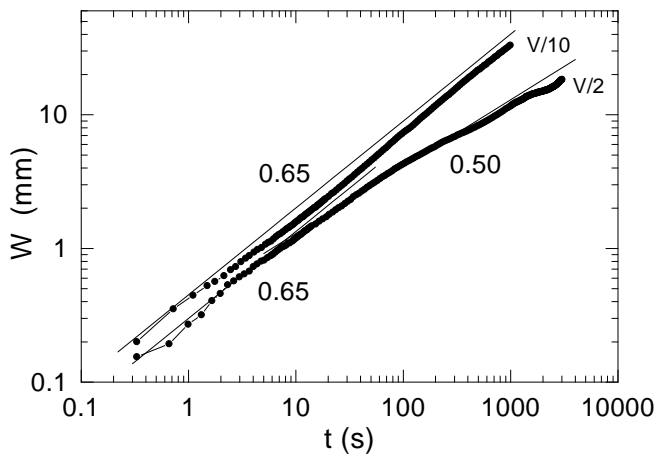} \vspace{0.5cm}
\caption{log-log plot of $W(t)$ for experiments at low velocity with disorder T 1.50 mm. The
end of the experiments is the moment at which the interfaces pinch--off. The gap spacing is
$b=0.36$ mm. Each curve is an average over two disorder configurations, with one run per
disorder configuration.} \label{Fig:low-v}
\end{center}
\end{figure}

At low injection rates, the external velocity $v$ can be tuned appropriately to have an
initial regime in which the oil advances over copper tracks and recedes over fiberglass
channels, followed by a regime in which the oil advances at all points of the interface. This
leads to two scaling regimes for $W(l,t)$, at short times with a growth exponent $\simeq
0.65$, which corresponds to a regime dominated by strong capillary forces, and at longer times
with a growth exponent $\simeq 0.50$, which corresponds to a regime where capillary and
viscous forces are better balanced. We have illustrated this behavior by performing
experiments at low $v$. In particular, we have studied velocities $V/2$ and $V/10$ using a gap
spacing $b=0.36$ mm and disorder T 1.50. The results are presented in Fig.\ \ref{Fig:low-v}.
The difficulty of these experiments is that the interface pinches--off before it reaches
saturation, so we can investigate only the growth regime. For $V/2$ we observe the two power
law regimes described before. For $V/10$, only the first regime can be observed before the
interface breaks. These results are qualitatively in agreement with the analysis of
Hern\'andez-Machado {\it et al.} \cite{Aurora-EPL-01}, who predicted two scaling regimes,
$\beta_1 = 5/6$ at short times (surface tension in the plane dominant) and $\beta_2 = 1/2$ at
long times (viscous pressure dominant). However, the presence of anomalous scaling and the
complicated role of capillary forces in our experiments makes it difficult a direct comparison
with the predictions of Hern\'andez-Machado {\it et al.}.

\begin{figure}
\begin{center}
\includegraphics[width=8.6cm]{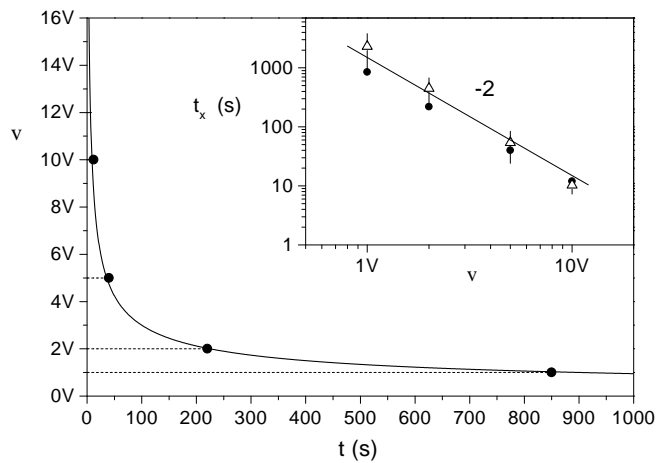} \vspace{0.5cm}
\caption{Prediction of the dependency of $t_{\times}$ with $v$. Main plot: for a given average
velocity $v$, the system saturates when the average velocity on copper tracks decays to $v$.
Inset: a log-log plot of the $t_{\times}$ obtained by the previous method (filled dots) as a
function of $v$ gives a straight line with slope $-2$. The triangles are the experimental
results for T 1.50 and $b=0.36$ mm.} \label{Fig:washburn}
\end{center}
\end{figure}

Another result than can be obtained from Eq.\ (\ref{Eq:Washburn-v}) is a prediction for the
saturation time $t_{\times}$ as a function of the average interface velocity $v$. We can
estimate $t_{\times}$ for a given $v$ and gap spacing $b$ by considering that saturation takes
place when all tracks reach the nominal velocity $v$. This is not strictly exact due to the
coupling between neighboring tracks, but it gives a good estimate of the dependency
$t_{\times}(v)$, as shown in Fig.\ \ref{Fig:washburn}. For a given experiment with velocity
$v_i < v_M$ the saturation takes place when the function $v_M \; t^{-1/2}$ crosses the nominal
velocity. Thus, as shown in the inset of the figure, $t_{\times}$ is expected to scale as
$t_{\times} \sim v^{-2}$. This result is confirmed by our experiments with disorder T 1.50 and
$b=0.36$ mm (open triangles in the inset).

\subsection{Experiments with a regular modulation of the gap spacing}\label{Subsec:Regular}

\begin{figure}
\begin{center}
\includegraphics[width=8.6cm]{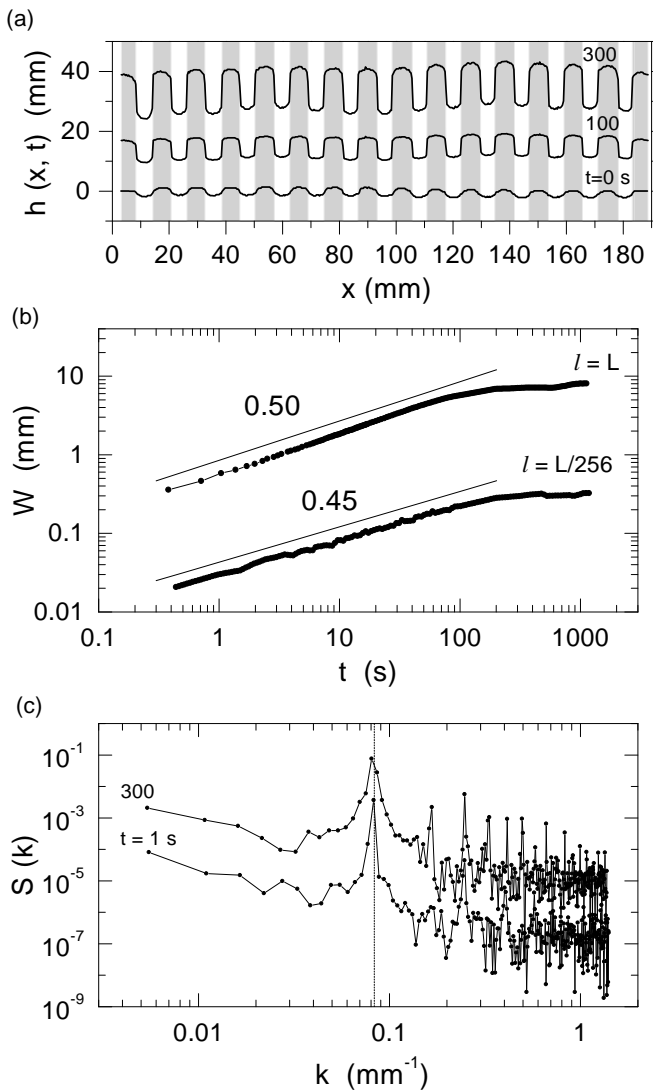} \vspace{0.5cm}
\caption{Experiments with a regular modulation of the gap spacing, using tracks 6 mm wide. The
experimental parameters are $v=2V=0.08$ mm/s and $b=0.36$ mm. (a) Temporal evolution of the
interfaces. (b) $W(l,t)$ plots for global ($l=L$) and local ($l=L/256$) scales. (c) Power
spectrum at saturation ($t=300$ s). The vertical line indicates the $k$ corresponding to the
spatial periodicity of the regular pattern.} \label{Fig:regular}
\end{center}
\end{figure}

In order to analyze the importance of the coupling between neighboring tracks, we have
performed experiments with a regular modulation of the gap spacing, consisting on copper
tracks $6$ mm wide separated by fiberglass channels $6$ mm wide. Interface velocity and
gap spacing are set to $v=0.08$ mm/s and $b=0.36$ mm respectively. The regular modulation
and the temporal evolution of the resulting interfaces are presented in Fig.\
\ref{Fig:regular}(a). Due to the symmetry of the regular pattern there is no growing
correlation length along the direction of the interface. For this reason, a dynamic
scaling of the interfacial fluctuations cannot even be defined. The growth of the
interfacial width is due to the independent, equivalent growth of interface fingers over
copper tracks, which follow the $t^{-1/2}$ behavior (Eq.\ \ref{Eq:Washburn-v}). The
system saturates when the fingers reach the nominal velocity. This is shown in Fig.\
\ref{Fig:regular}(b), which shows the evolution of $W(l,t)$ for the global ($l=L$) and
local ($l=L/256$) scales. The figure reveals that both scales grow in the same way. The
absence of dynamic scaling is also confirmed by the power spectrum, shown in Fig.\
\ref{Fig:regular}(c). We obtain a nearly flat spectrum, with a large peak at the value of
$k$ that corresponds to the spatial periodicity of the underlying regular pattern. We
also observe a vertical shift of the spectra at different times, caused by the
progressive elongation of the oil fingers over copper tracks during growth.

The previous experiments with a regular modulation of the gap spacing illustrate the need
of a growing correlation length along the interface to reach a dynamic scaling regime.
Similarly to these experiments, the lateral growth of the correlation length can also be
inhibited when the modulation of the gap spacing is random, provided that the
experimental parameters are such that the motion over tracks is not coupled and, on each
track, follows Eq.\ (\ref{Eq:Washburn-v}). Thus, for very small gap spacings ($b \leq
0.16$ mm) and $v \geq V$ we observe a growth of the interfacial width with exponent
$\simeq 0.5$ for both global and local scales. Similarly, in the experiments at $Q=0$
with $b=0.16$ mm the same exponent $\simeq 0.65$ is measured for both global and local
scales. The power spectrum of this case also demonstrates the lack of dynamic scaling, as
can be seen in the top plot of Fig.\ \ref{Fig:q0-w-t}(b). When the modulation of the gap
spacing is regular, the transition to saturation is sharp due to the fact that all tracks
have the same width, and thus all the oil fingers reach the nominal velocity at the same
time. For a random pattern the transition to saturation is smooth, and does not take
place strictly until the widest copper tracks (those that have the highest value of
$A^i_{+}$) reach the nominal velocity. The plot for $b=0.16$ mm of Fig.\
\ref{Fig:local-b} illustrates this behavior.

\section{Discussion and conclusions}\label{Sec:Conclusions}

\begin{figure}
\begin{center}
\includegraphics[width=8.6cm]{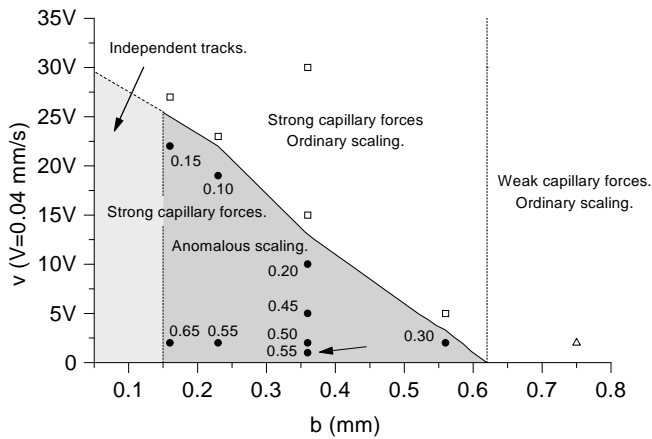} \vspace{0.5cm}
\caption{Phase diagram for the anomalous scaling. The symbols indicate the parameters explored
experimentally. The numbers beside the symbols in the anomalous scaling region give the value
of $2 \theta / z $. The solid line represents the curve $v=v_{M}(b)$. The arrow indicates the
particular experiment of Figs.\ \ref{Fig:T-V} and \ref{Fig:collapse}.} \label{Fig:diagrama}
\end{center}
\end{figure}

Our experimental results show that two ingredients are necessary to observe anomalous scaling
in a disordered Hele--Shaw cell: i) strong destabilizing capillary forces, sufficiently
persistent in the direction of growth, and ii) coupling in the motion between neighboring
tracks. These two ingredients give rise to sharp slopes at the track edges together with a
growing correlation length along the interface. This scenario is represented in the phase
diagram of Fig.\ \ref{Fig:diagrama}, which shows the different regions of scaling explored
experimentally. Each point of the diagram is an average over 6 experiments (3 disorder
configurations and 2 runs per disorder configuration). The dark grey area represents the
region where an anomalous scaling has been identified. We have indicated the value of $2
\theta /z$, which is the value needed to collapse vertically the spectral curves and,
therefore, indicates the degree of anomaly. For $0.15 \leq b \leq 0.60$ mm, and for driving
velocities below $v_M(b)$, the capillary forces are strong enough to allow an initial
acceleration and subsequent deceleration of the oil fingers over copper tracks, but
sufficiently weak to allow coupling between neighboring tracks. For $v
> v_M$, the viscous forces become dominant, and the anomalous scaling is unobservable.
For $b > 0.60$ mm, the capillary forces are not sufficiently strong and we have ordinary
scaling for any velocity. Finally, for $b \leq 0.15$ mm (light grey area in Fig.\
\ref{Fig:diagrama}), the capillary forces are dominant, and the oil fingers over copper tracks
and fiberglass channels have independent motion. In this situation a dynamic scaling scenario
is not adequate because there is no growing correlation length along the interface. Although
$W(l,t)$ grows as a power law of time, this is a consequence of the independent behavior on
each track given by Eq.\ (\ref{Eq:Washburn}). Although it is not possible to determine the
scaling exponents, still is possible to perform a vertical collapse of the spectral curves
because only the {\it difference} between $\alpha$ and $\alpha_{loc}$ (i.e. $\theta$) is
needed. This allows to give the values of $2 \theta / z$ for $b=0.16$ mm.

In conclusion, we have reported experiments of anomalous kinetic roughening in Hele--Shaw
flows with quenched disorder. We have studied the displacement of an oil--air interface for
different strength of the capillary forces, flow velocity, and disorder configurations. We
have observed that the scaling of the interfacial fluctuations follows the intrinsic anomalous
scaling ansatz in conditions of sufficiently strong capillary forces, low velocities, and a
persistent disorder in the growth direction. A random distribution of tracks has been used as
representative of a persistent disorder. A different scaling for the global and local
interfacial fluctuations has been measured, with exponents $\beta \simeq 0.50$, $\beta^{\ast}
\simeq 0.25$, $\alpha \simeq 1.0$, $\alpha_{loc} \simeq 0.5$, and $z \simeq 2$. The presence
of anomalous scaling has been confirmed by the existence of multiscaling. When capillary
forces become dominant or when a regular track pattern is used, however, there is no growing
correlation length along the interface, and therefore no dynamic scaling. A detailed
investigation of the interface dynamics at the scale of the disorder shows that the anomaly is
a consequence of the different velocities over copper tracks and fiberglass channels plus the
coupled motion between neighboring tracks. The resulting dynamics gives interface profiles
characterized by high slopes at the edges of the copper tracks. These slopes follow an
anomalous L\'evy distribution characteristic of systems which display anomalous scaling.

\section{Acknowledgements}
We are grateful to M. A. Rodr\'{\i}guez, J. Ramasco, and R. Cuerno for fruitful discussions.
The research has received financial support from the Direcci\'on General de Investigaci\'on
(MCT, Spain), project BFM2000-0628-C03-01. J. O. acknowledges the Generalitat de Catalunya for
additional financial support. J. S. was supported by a fellowship of the DGI (MCT, Spain)  in
the period 1998-2001.

\end{document}